\documentclass[12pt]{article}
\usepackage{graphicx}
\usepackage{calc}
\usepackage{hyperref}
\hypersetup{breaklinks,colorlinks=true}
\usepackage{breakurl}
\newcommand{\dpr}{doi: \href{http://dx.doi.org/\dbi}{\nolinkurl{\dbi}}}
\newcommand{\upr}{url: \url{\URL}}

\newcommand\pubnumber{}
\newcommand\pubdate{}

\newcommand{\mathtf}{\mathrm}

\newcommand{\tev}{\mathtf{TeV}}
\newcommand{\gev}{\mathtf{GeV}}

\newcommand{\fb}{\mathtf{fb}}

\newcommand{\ps}{\mathtf{ps}}

\newcommand{\sm}{\mathtf{SM}}
\newcommand{\rad}{\mathtf{rad}}
\newcommand{\ifb}{\fb^{-1}}

\newcommand{\ips}{\ps^{-1}}

\newcommand{\lumi}{{\cal L}}

\newcommand{\sqrs}{\sqrt{s}}
\newcommand{\eold}{\sqrs =    7~\tev}
\newcommand{\enew}{\sqrs =    8~\tev}

\newcommand{\lold}{\lumi \sim    5~\ifb}

\newcommand{\lnew}{\lumi \sim   20~\ifb}

\newcommand{\angth}{\theta}

\newcommand{\jpphi}{\varphi}
\newcommand{\tht}{\angth_T}
\newcommand{\phs}{\phi_s}
\newcommand{\gs}{\Gamma_s}
\newcommand{\dgs}{\Delta \gs}
\newcommand{\pht}{\jpphi_T}
\newcommand{\pst}{\psi_T}

\newcommand{\ssp}{\sin \pht}
\newcommand{\sss}{\sin \pst}

\newcommand{\csp}{\cos \pht}

\newcommand{\sdt}{\sin 2\tht}
\newcommand{\sdp}{\sin 2\pht}
\newcommand{\sds}{\sin 2\pst}

\newcommand{\sqt}{\sin^2\tht}
\newcommand{\sqp}{\sin^2\pht}
\newcommand{\sqs}{\sin^2\pst}

\newcommand{\cqp}{\cos^2\pht}
\newcommand{\cqs}{\cos^2\pst}
\newcommand{\fst}{\frac{1}{\sqrt{2}}}
\newcommand{\ftt}{\frac{2}{3}}
\newcommand{\fts}{\frac{1}{3}\sqrt{6}}
\newcommand{\ffs}{\frac{4}{3}\sqrt{3}}

\newcommand{\dcpv}{|\lambda|}
\newcommand{\dcpq}{\dcpv^2}
\newcommand{\ab}{A_\bot}
\newcommand{\az}{A_0}
\newcommand{\al}{A_\|}
\newcommand{\as}{A_S}
\newcommand{\aab}{|\ab|}
\newcommand{\aaz}{|\az|}
\newcommand{\aal}{|\al|}
\newcommand{\aas}{|\as|}
\newcommand{\db}{\delta_\bot}
\newcommand{\dz}{\delta_0}
\newcommand{\dl}{\delta_\|}
\newcommand{\ds}{\delta_S}
\newcommand{\dd}{\delta_{S\bot}}

\newcommand{\jpsi}{{J/\psi}}

\newcommand{\bp}{B^+}
\newcommand{\bm}{B^-}
\newcommand{\bz}{B^0}
\newcommand{\bd}{B^0_d}
\newcommand{\bs}{B^0_s}

\newcommand{\bb}{\bar{B}^0_s}

\newcommand{\bsjphi}{\bs \rightarrow \jpsi \phi}

\newcommand{\dbi}{}
\newcommand{\URL}{}

\newcommand{\typeaut}{}

\newcommand{\typetit}{}
\newcommand{\typeref}{}

\newcommand{\typedoi}{}
\newcommand{\typeurl}{}

\def\padova{Dipartimento di Fisica e Astronomia\\
Universit\`a di Padova and INFN, I-35131 Padova, ITALY}
\def\Title#1{\begin{center} {\Large #1 } \end{center}}
\def\Author#1{\begin{center}{ \sc #1} \end{center}}
\def\Address#1{\begin{center}{ \it #1} \end{center}}

\newcommand\pubblock{\rightline{\begin{tabular}{l} \pubnumber\\
         \pubdate  \end{tabular}}}
\newenvironment{Abstract}{\begin{quotation}  }{\end{quotation}}
\newenvironment{Presented}{\begin{quotation} \begin{center} 
             PRESENTED AT\end{center}\bigskip 
      \begin{center}\begin{large}}{\end{large}\end{center} \end{quotation}}

\def\beq{\begin{equation}}
\def\eeq#1{\label{#1}\end{equation}}
\def\eeqn{\end{equation}}

\def\beqa{\begin{eqnarray}}
\def\eeqa#1{\label{#1}\end{eqnarray}}
\def\eeqan{\end{eqnarray}}

\let\bar=\overbar

\def\Dslash{\not{\hbox{\kern-4pt $D$}}}
\def\dslash{\not{\hbox{\kern-2pt $\del$}}}

\def\msb{{\bar{\ssstyle M \kern -1pt S}}}

\def\Journal#1#2#3#4{{#1} {\bf #2}, #3 (#4)}

\def\PRD{{\typeref Phys. Rev.} D}

\begin{document}
\begin{titlepage}
\pubblock

\vfill
\Title{Measurement of the CP violating phase $\phi_s$ \\
with CMS and ATLAS}
\vfill
\Author{Paolo Ronchese}
\Address{\padova}
\vfill
\begin{Abstract}
The phase $\phi_s$ is the key parameter for the CP-violation of the 
$\bs$-$\bar{\bs}$ system. An angular and proper decay time analysis is 
applied to the $\bsjphi$ events. Using a data sample collected by the CMS and 
ATLAS experiments in LHC Run1, the $\bs$ signal candidates are reconstructed 
and are used to extract the phase $\phi_s$. We present the latest update on 
the results in this decay channel. 

\end{Abstract}
\vfill
\begin{Presented}
Twelfth Conference on the Intersections \\
of Particle and Nuclear Physics\\
Vail, Colorado (USA), May 19-24, 2015\rule{0mm}{15pt}
\end{Presented}
\vfill
\end{titlepage}
\def\thefootnote{\fnsymbol{footnote}}
\setcounter{footnote}{0}

\section{Introduction}

The motivations to study CP violation in $\bs$ decay to $\jpsi \phi$ is the 
look for indirect evidence, or constraints, of new physics beyond the 
standard model.

In the $\bsjphi$ decay the flavoured initial state is an admixture of two 
mass eigenstates $\bz_L$ and $\bz_H$, while the final state is 
unflavoured, so an interference arises between the direct and mixing-mediated 
decays. 

The mixing process is described by a box diagram, 
and it is affected by the presence of new particles circulating in the loop. 
In particular, the decay $\bsjphi$ is theoretically clean, 
having a vanishing phase in the quark-level decay $b \rightarrow c\bar{c}s$. 
On the other side this decay has important differences when compared to 
the corresponding decay $\bd \rightarrow \jpsi K^0_s$.
As for $\bd$ the initial state is actually an admixture of two mass 
eigenstates with different masses and lifetimes, but compared to $\bd$
the mass difference is much bigger so the oscillation is much faster, and 
the width difference is no more negligible. Another important difference, 
the final state does not have a definite CP, being an admixture of odd 
and even CP eigenstates, that must be disentangled performing a 
time-dependent angular analysis. 

\begin{figure}[htb]
  \centering
  \includegraphics[height=115pt]{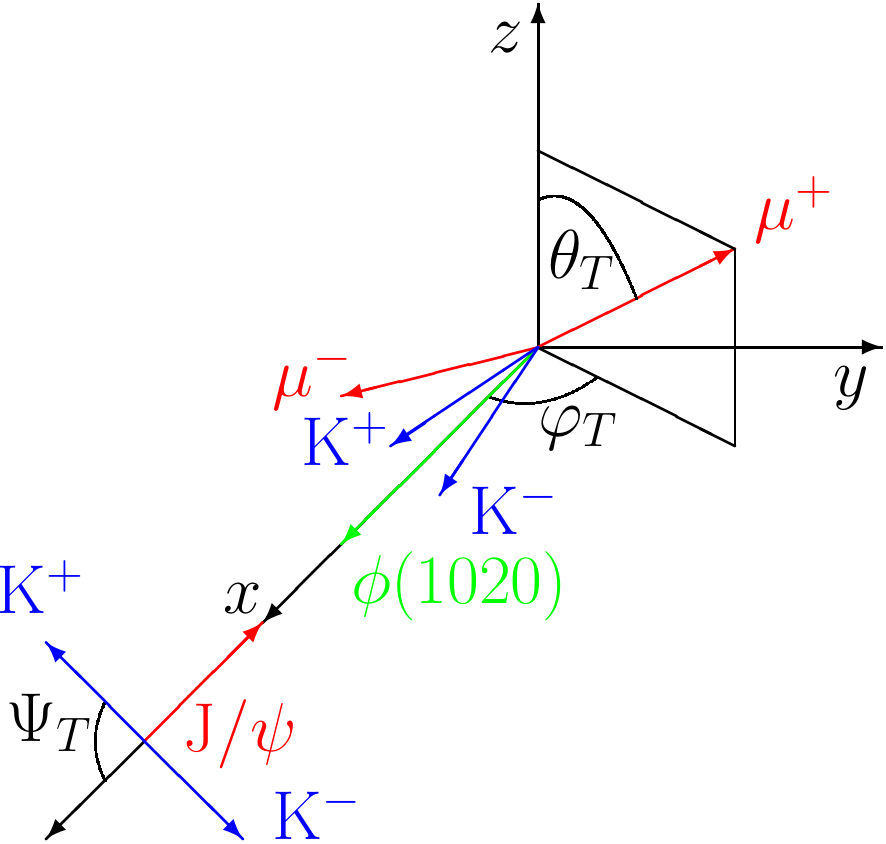}
  \caption{Decay angles in $\bsjphi$.}
  \label{fig:angle}
\end{figure}

The interference phase in the standard model is predicted to be 
$\phs \simeq -2\beta_s$, where 
$\beta_s = \mathrm{arg}(-(V_{ts}V_{tb}^*)/(V_{cs}V_{cb}^*))$. 
The latest prediction~\cite{ref:bjp2b} for $\beta_s$ is
\begin{displaymath}
  2\beta_{s(\sm)} = 0.0363^{+0.0016}_{-0.0015}~\rad.
\end{displaymath}

Very small deviations from 
the standard model expectation are also predicted by new physics models with 
minimal flavour violation, $|\phs| < 0.05$, so that any observation of a 
large CP violation phase would rule out those models~\cite{ref:bjpmv}. 
The analysis allows also a determination of the decay width difference, 
predicted by the standard model to be 
\begin{displaymath}
  \Delta\Gamma_{s(\sm)} = 0.087 \pm 0.021~\ips,
\end{displaymath}
but this quantity is expected having a reduced sensitivity to new physics 
processes~\cite{ref:bjpwd}. 

\section{Data samples and selections}

The results obtained from the analysis of data collected in 2011 at $\eold$ 
by ATLAS, corresponding to an integrated luminosity $\lold$, 
%$\lumi = 4.7~\ifb$, 
and in 2012 at $\enew$ by CMS, corresponding to an integrated luminosity 
$\lnew$, will be shown in the following.

Dedicated triggers have been developed for the analyses, requiring the 
presence of two muons forming a secondary vertex displaced from the 
primary interaction point, to achieve a sustainable trigger rate when 
collecting data at the very high luminosities provided by LHC.
%Dedicated triggers have been developed for the analyses, to achieve a 
%sustainable trigger rate when collecting data at the very high luminosities 
%provided by LHC, requiring the presence of two muons forming a secondary 
%vertex displaced from the primary interaction point.
Events have been selected requiring two 
opposite charge muons and two other opposite charge tracks assumed to be 
kaons, forming a $\jpsi$ and a $\phi$ candidate respectively, with a common 
vertex.

\section{Data analysis}

The differential decay width is expressed as the sum of 10 functions 
of the time and the angles $\Theta = \tht , \pht , \pst$, as defined in 
fig.\ref{fig:angle}:

\begin{displaymath}
  \frac{d^4\Gamma(\bs(t))}{d\Theta dt} = f(\Theta, \alpha, ct) \propto
  \sum_{i=1}^{10} O_i(\alpha, ct) \cdot g_i(\Theta)
\end{displaymath}

\begin{displaymath}
  \begin{array}{l}
  O_i(\alpha, ct) = N_i e^{-t/\tau}\left[
    a_i\cosh\left(\frac{1}{2}\dgs ct\right) +
    b_i\sinh\left(\frac{1}{2}\dgs ct\right) \right.\\
    \rule{97pt}{0mm}\pm\left.
    c_i\cos \left(           \Delta     m_s ct\right) \pm
    d_i\sin \left(           \Delta     m_s ct\right)
    \makebox[0mm][l]{\rule{0pt}{\heightof{$\frac{1}{2}$}}}\right]
  \end{array}
\end{displaymath}

{
  \newlength{\icl}
  \newlength{\gcl}
  \newlength{\ncl}
  \newlength{\acl}
  \newlength{\bcl}
  \newlength{\ccl}
  \newlength{\dcl}
  \setlength{\icl}{7pt}
  \setlength{\gcl}{92pt}
  \setlength{\ncl}{24pt}
  \setlength{\acl}{44pt}
  \setlength{\bcl}{44pt}
  \setlength{\ccl}{44pt}
  \setlength{\dcl}{44pt}
  \tiny
%
%  \setlength{\icl}{6pt}
%  \setlength{\gcl}{88pt}
%  \setlength{\ncl}{20pt}
%  \setlength{\acl}{37pt}
%  \setlength{\bcl}{40pt}
%  \setlength{\ccl}{35pt}
%  \setlength{\dcl}{40pt}
%  \tiny
%
%  \setlength{\icl}{8pt}
%  \setlength{\gcl}{115pt}
%  \setlength{\ncl}{30pt}
%  \setlength{\acl}{53pt}
%  \setlength{\bcl}{60pt}
%  \setlength{\ccl}{53pt}
%  \setlength{\dcl}{55pt}
%  \scriptsize
    \hspace*{-1mm}
    \begin{center}
    \begin{tabular}{r|c|ccccc} 
      \makebox[\icl][c]{$i$} & $g_i(\tht,\pht,\pst)$
    & $N_i$ & $a_i$ & $b_i$ & $c_i$ & $d_i$ \\ \hline
      \makebox[\icl][r]{1} & \makebox[\gcl][c]{$2\cqs(1-\sqt\cqp)$}
    & \makebox[\ncl][c]{$\aaz^2  $}
    & \makebox[\acl][c]{$ 1             $}
    & \makebox[\bcl][c]{$ D             $}
    & \makebox[\ccl][c]{$ C             $}
    & \makebox[\dcl][c]{$-S             $} \\
      \makebox[\icl][r]{2} & \makebox[\gcl][c]{$ \sqs(1-\sqt\sqp)$}
    & \makebox[\ncl][c]{$\aal^2  $}
    & \makebox[\acl][c]{$ 1             $}
    & \makebox[\bcl][c]{$ D             $}
    & \makebox[\ccl][c]{$ C             $}
    & \makebox[\dcl][c]{$-S             $} \\
      \makebox[\icl][r]{3} & \makebox[\gcl][c]{$ \sqs   \sqt     $}
    & \makebox[\ncl][c]{$\aab^2  $}
    & \makebox[\acl][c]{$ 1             $}
    & \makebox[\bcl][c]{$-D             $}
    & \makebox[\ccl][c]{$ C             $}
    & \makebox[\dcl][c]{$ S             $} \\
      \makebox[\icl][r]{4} & \makebox[\gcl][c]{$-\sqs\sdt\ssp    $}
    & \makebox[\ncl][c]{$\aaz\aab$}
    & \makebox[\acl][c]{$ C\sin(\db-\dl)$}
    & \makebox[\bcl][c]{$ S\cos(\db-\dl)$}
    & \makebox[\ccl][c]{$  \sin(\db-\dl)$}
    & \makebox[\dcl][c]{$ D\cos(\db-\dl)$} \\
      \makebox[\icl][r]{5} & \makebox[\gcl][c]{$\fst\sds\sqt\sdp $}
    & \makebox[\ncl][c]{$\aaz\aal$}
    & \makebox[\acl][c]{$  \cos(\dl-\dz)$}
    & \makebox[\bcl][c]{$ D\cos(\dl-\dz)$}
    & \makebox[\ccl][c]{$ C\cos(\dl-\dz)$}
    & \makebox[\dcl][c]{$-S\cos(\dl-\dz)$} \\
      \makebox[\icl][r]{6} & \makebox[\gcl][c]{$\fst\sds\sdt\ssp $}
    & \makebox[\ncl][c]{$\aaz\aab$}
    & \makebox[\acl][c]{$ C\sin(\db-\dz)$}
    & \makebox[\bcl][c]{$ S\cos(\db-\dz)$}
    & \makebox[\ccl][c]{$  \sin(\db-\dz)$}
    & \makebox[\dcl][c]{$ D\cos(\db-\dz)$} \\
      \makebox[\icl][r]{7} & \makebox[\gcl][c]{$\ftt(1-\sqt\cqp)$}
    & \makebox[\ncl][c]{$\aas^2$}
    & \makebox[\acl][c]{$ 1             $}
    & \makebox[\bcl][c]{$-D             $}
    & \makebox[\ccl][c]{$ C             $}
    & \makebox[\dcl][c]{$ S             $} \\
      \makebox[\icl][r]{8} & \makebox[\gcl][c]{$\fts\sss\sqt\sdp$}
    & \makebox[\ncl][c]{$\aas\aal$}
    & \makebox[\acl][c]{$ C\cos(\dl-\ds)$}
    & \makebox[\bcl][c]{$ S\sin(\dl-\ds)$}
    & \makebox[\ccl][c]{$  \cos(\dl-\ds)$}
    & \makebox[\dcl][c]{$ D\sin(\dl-\ds)$} \\
      \makebox[\icl][r]{9} & \makebox[\gcl][c]{$\fts\sss\sdt\csp$}
    & \makebox[\ncl][c]{$\aas\aab$}
    & \makebox[\acl][c]{$  \sin(\db-\ds)$}
    & \makebox[\bcl][c]{$-D\sin(\db-\ds)$}
    & \makebox[\ccl][c]{$ C\sin(\db-\ds)$}
    & \makebox[\dcl][c]{$ S\sin(\db-\ds)$} \\
      \makebox[\icl][r]{10} & \makebox[\gcl][c]{$\ffs(1-\sqt\cqp)$}
    & \makebox[\ncl][c]{$\aas\aaz$}
    & \makebox[\acl][c]{$ C\cos(\dz-\ds)$}
    & \makebox[\bcl][c]{$ S\sin(\dz-\ds)$}
    & \makebox[\ccl][c]{$  \cos(\dz-\ds)$}
    & \makebox[\dcl][c]{$ D\sin(\dz-\ds)$} \\ \hline
    \end{tabular}
    \end{center}
}

\begin{displaymath}
    C =  \frac{1-\dcpq}{1+\dcpq}
    \rule{10mm}{0mm}
    S = -\frac{2\dcpv\sin\phs}{1+\dcpq}
    \rule{10mm}{0mm}
    D = -\frac{2\dcpv\cos\phs}{1+\dcpq}.
\end{displaymath}

The amplitudes $\ab$, $\az$, $\al$, $\as$ correspond to the $P$-wave and 
$S$-wave components, with their phases $\db$, $\dz$, $\dl$, $\ds$; 
$\dcpv$ describes the direct CP violation.
In the expression the signs of $c_i$ and $d_i$ coefficients are positive 
or negative for the decay of an initial $\bs$ or $\bb$ respectively. 

The differential width depends only on the differences among the phases 
$\delta$, so in the fit $\dz = 0$ was assumed, and the difference $\dd$ 
between $\db$ and $\ds$ was fitted as an unique variable to reduce the 
correlation among the parameters. 
No direct violation was assumed in the measurement, therefore 
$\dcpv = 1$ was fixed.

The discrimination between the positive or negative initial flavour is 
obtained looking for a second $B$ produced in the event and inferring 
its flavour looking at the charge of its decay products; of course the 
charge-flavour 
correlation is diluted due to the presence of cascade decays and 
oscillations of the other $b$ hadron itself.

\begin{figure}[htb]
  \centering
  \includegraphics[height=36mm]{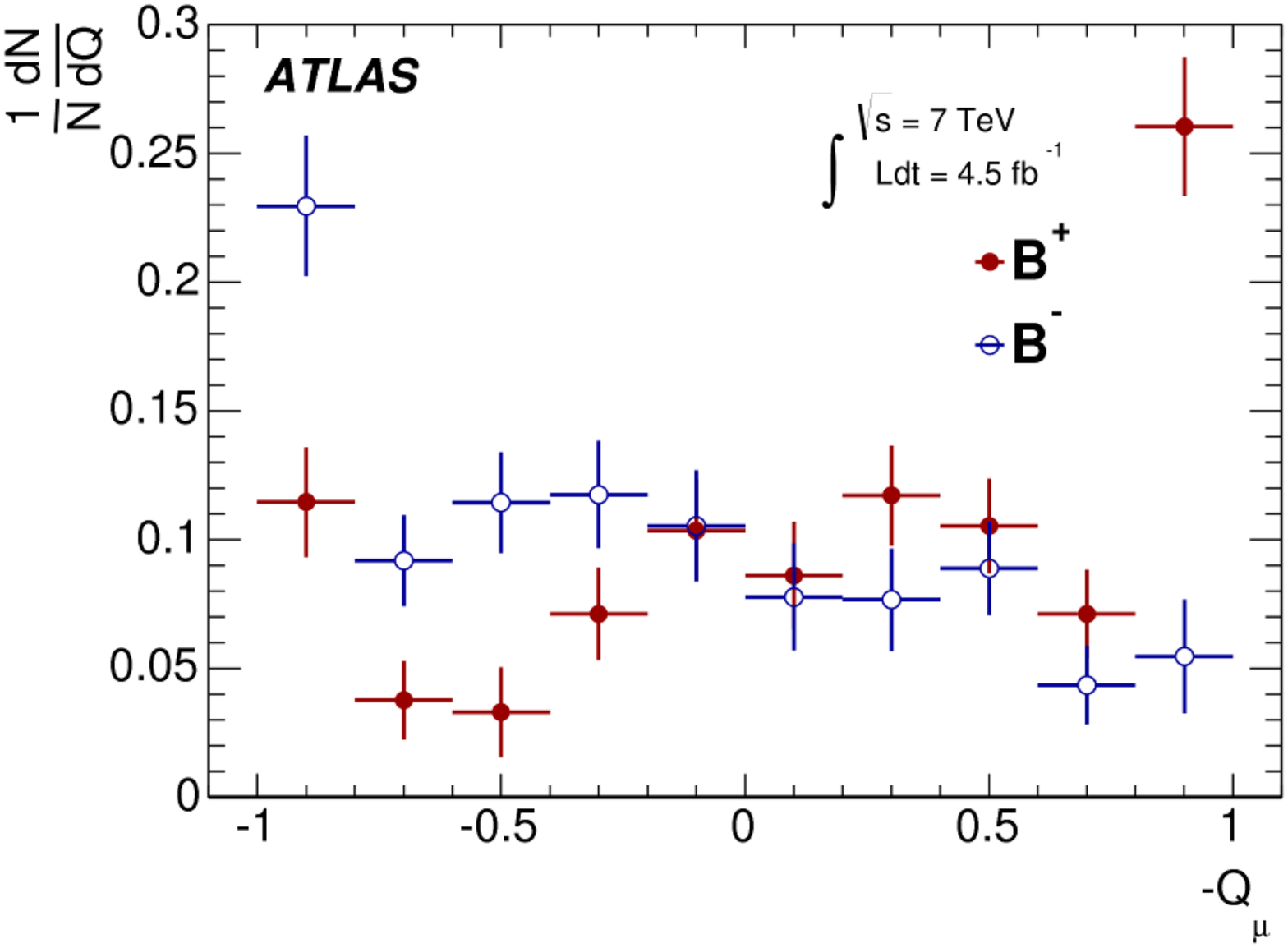}
  \includegraphics[height=36mm]{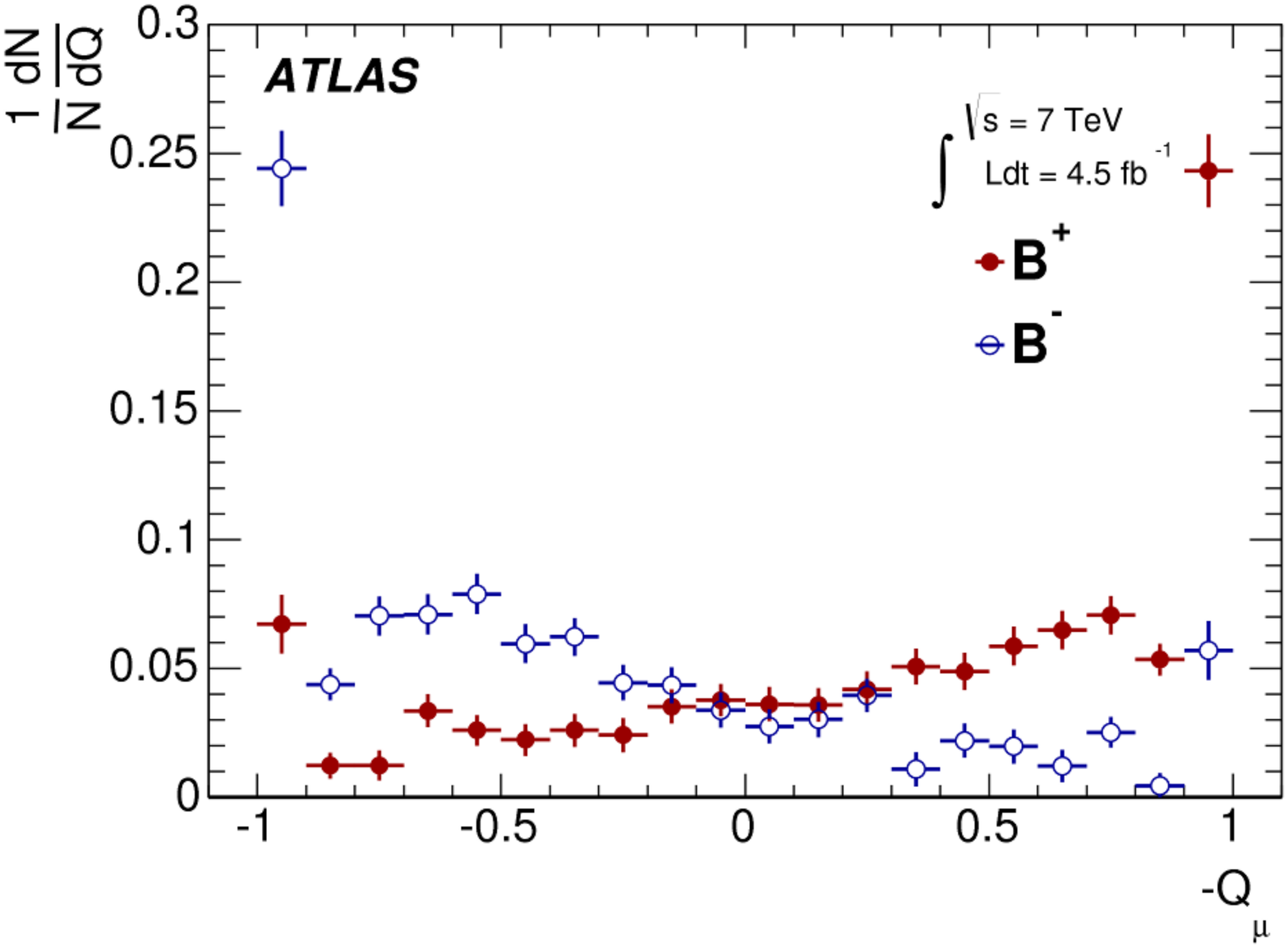}
  \includegraphics[height=36mm]{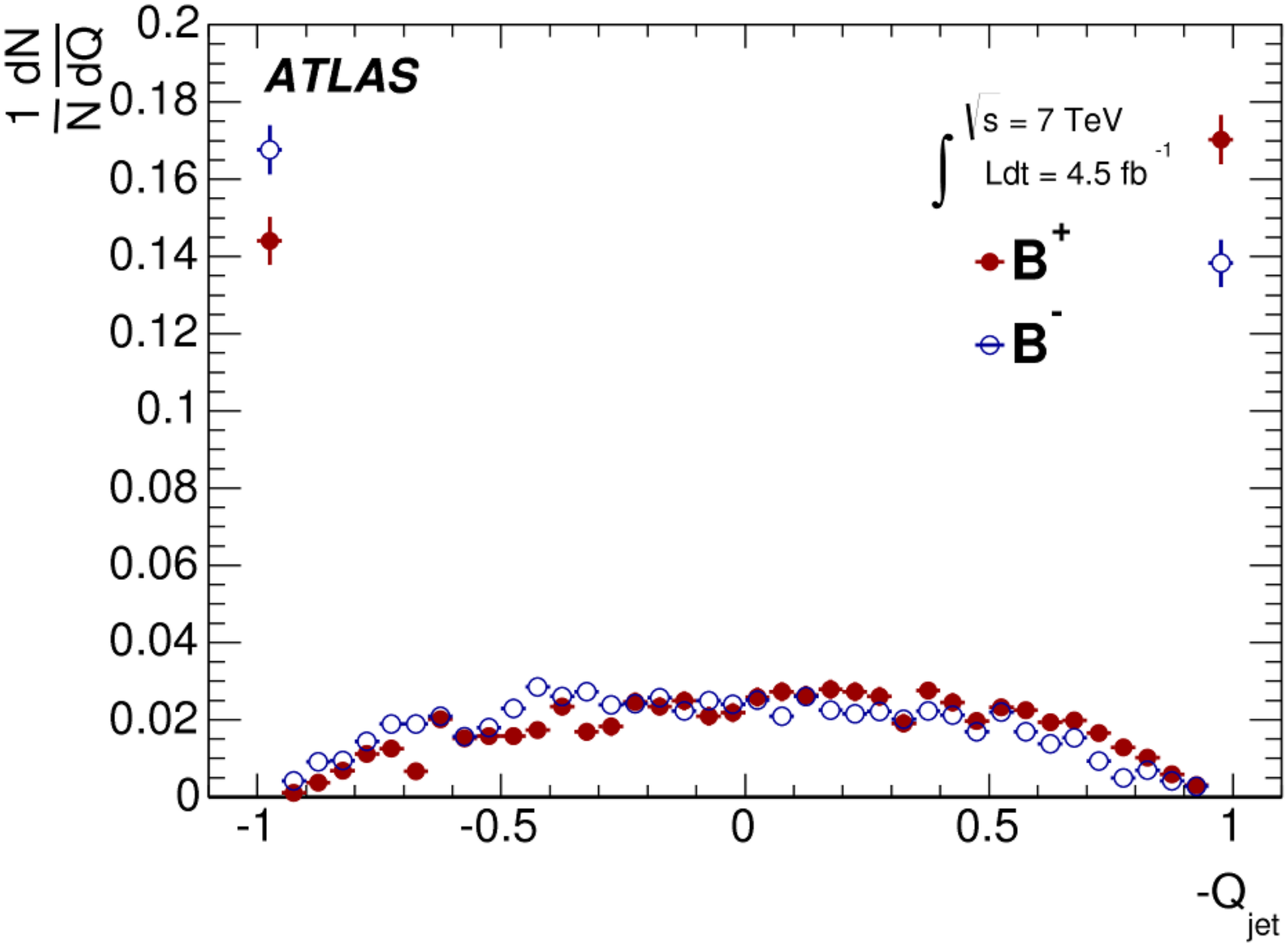}
  \caption{Opposite-side charge for $\bp$ and $\bm$, for segmented muons 
    (left), combined muons (middle) and jets (right) in ATLAS analysis.}
  \label{fig:atlft}
\end{figure}

In ATLAS analysis~\cite{ref:bjpat} of $\eold$ data the flavour was tagged 
looking at the charge of particles contained in a cone around a muon, 
assumed to 
come from the semileptonic decay of a $b$; muons were classified
according to their reconstruction class, combined or segment; if no muon 
was found the particles 
in a $b$-tagged jets were used. A ``cone charge'' was defined as 

\begin{displaymath}
    Q = \frac{\sum_i^{N_\mathrm{tracks}} q^i \cdot (p_{T,i})^j}
             {\sum_i^{N_\mathrm{tracks}}           (p_{T,i})^j}
\end{displaymath}

where $j = 1.1$ and the sum was performed over the reconstructed tracks 
within a cone size of $\Delta R = 0.5$. Charge distributions for $\bp$ 
and $\bm$ are shown in fig.\ref{fig:atlft} and tagging performances are 
summarized in tab.\ref{tab:atlft}

\begin{figure}[htb]
  \centering
  \includegraphics[height=36mm]{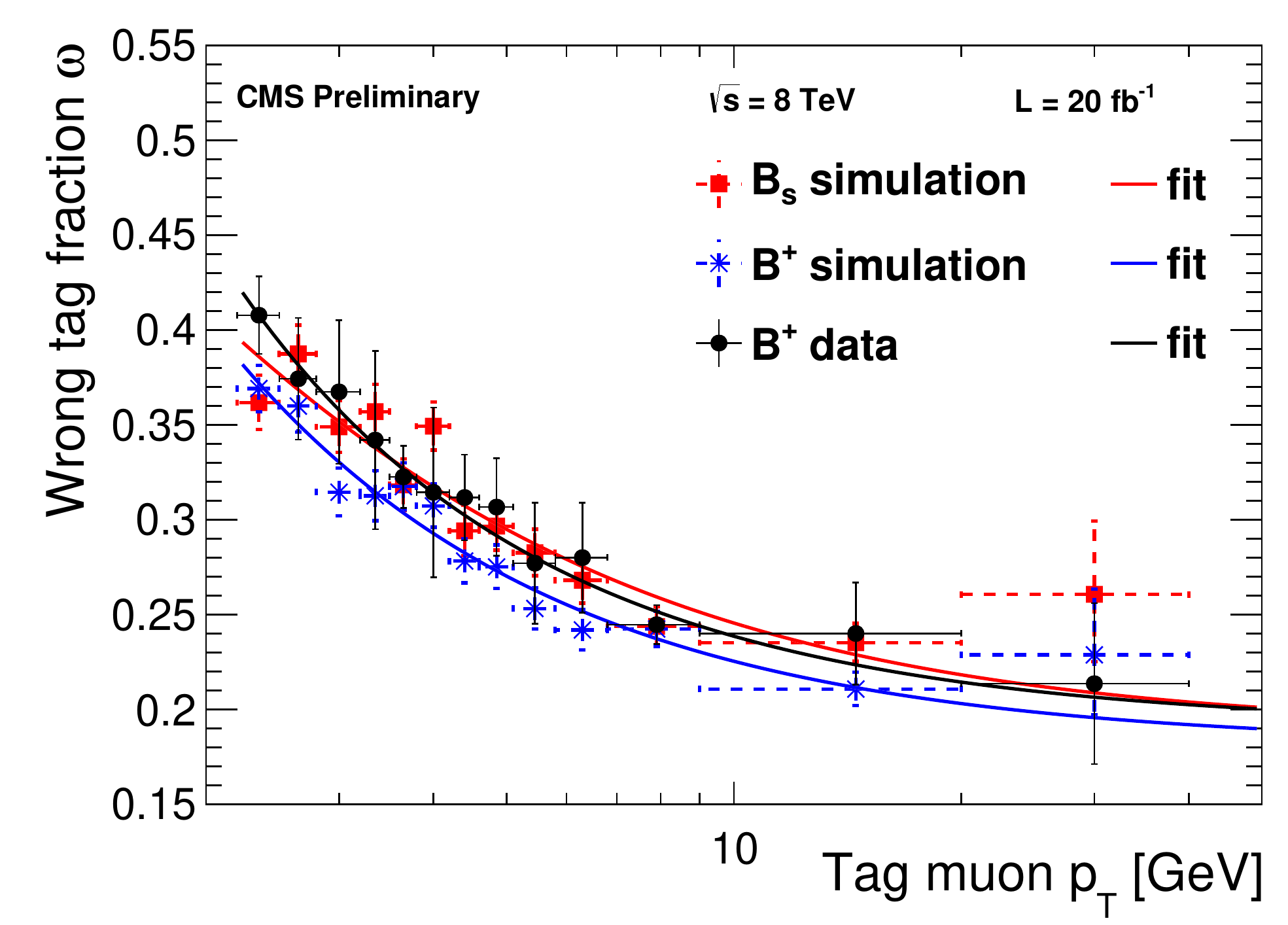}
  \includegraphics[height=36mm]{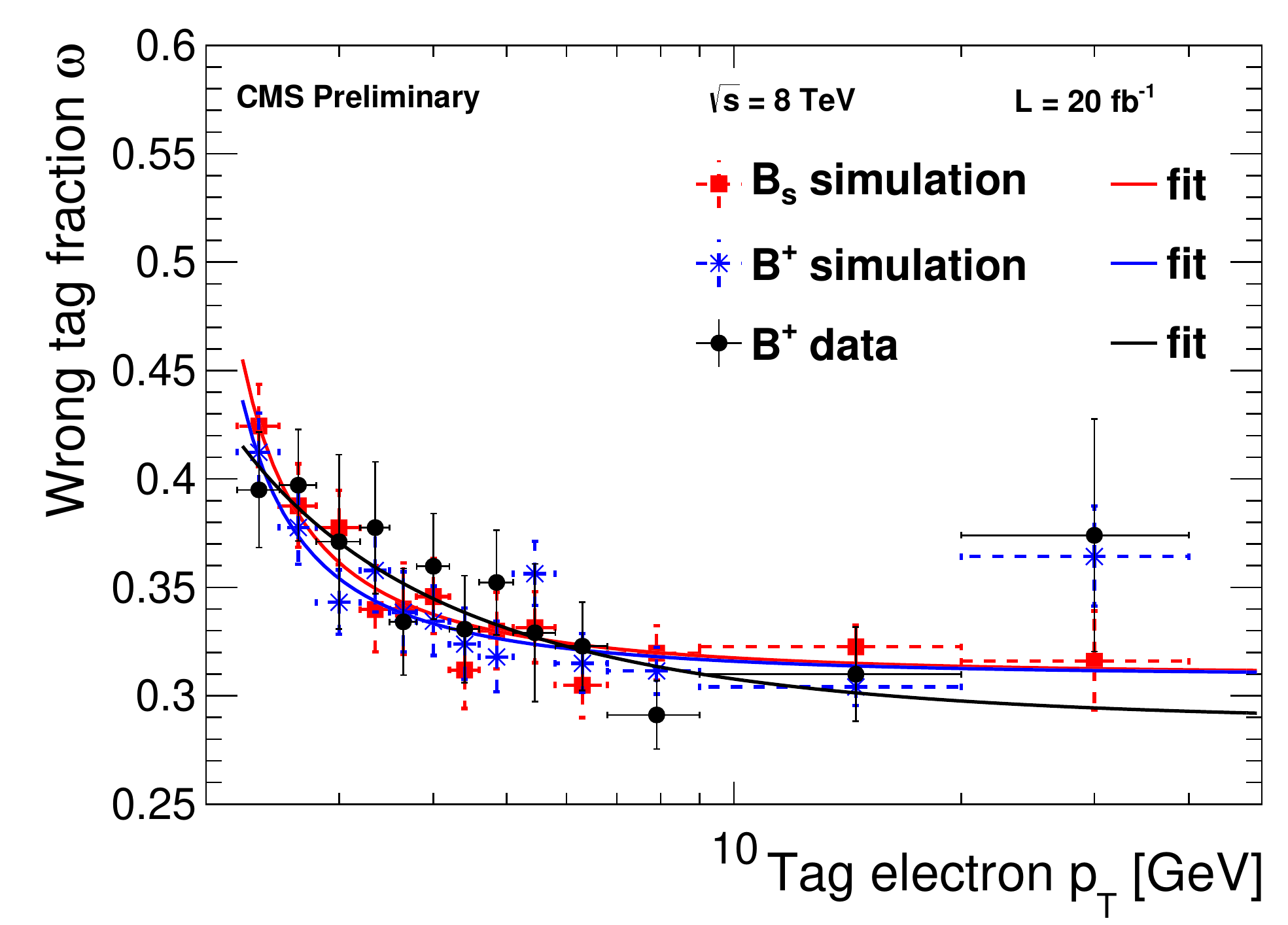}
  \caption{Mistag fraction versus transverse momentum for muons (left) 
    and electrons (right) in CMS analysis.}
  \label{fig:cmsft}
\end{figure}

In CMS analysis~\cite{ref:bjpcm} of $\enew$ data only semileptonic decays 
were used to 
tag the flavour, looking to both electrons and muons; in fig.\ref{fig:cmsft} 
the mistag fraction versus transverse momentum is shown.

The performance 
of the methods were measured with events containing the self-tagging 
decay $\bp \rightarrow \jpsi K^+$. The efficiency and tagging 
power of the algorithms are reported in 
tab.\ref{tab:atlft} for ATLAS and tab.\ref{tab:cmsft} for CMS. 
%tab.\ref{tab:bjpft}. 

\begin{table}[htb]
  \begin{center}
    \begin{tabular}{l|lll}
                           & \multicolumn{1}{c}{Muons (c)}
                           & \multicolumn{1}{c}{Muons (s)}
                           & \multicolumn{1}{c}{$b$-jet} \\ \hline
      Efficiency      [\%] & $3.37  \pm 0.04$
                           & $1.08  \pm 0.02$
                           & $27.7  \pm 0.1 $ \\
      Dilution        [\%] & $50.6  \pm 0.5 $
                           & $36.7  \pm 0.7 $
                           & $12.68 \pm 0.06$ \\
      Tagging power   [\%] & $0.86 \pm 0.04$
                           & $0.15 \pm 0.02$
                           & $0.45 \pm 0.03$ \\ \hline
    \end{tabular}
  \end{center}
  \caption{Performance of flavour tagging algorithm in ATLAS analysis; errors 
    are statistical only.}
  \label{tab:atlft}
\end{table}

\begin{table}[htb]
  \begin{center}
    \begin{tabular}{l|ll}
                           & \multicolumn{1}{c}{Muons}
                           & \multicolumn{1}{c}{Electrons} \\ \hline
      Efficiency      [\%] & $4.55 \pm 0.03 \pm 0.08$
                           & $3.26 \pm 0.02 \pm 0.01$ \\
      Mistag fraction [\%] & $30.7 \pm 0.4 \pm 0.7$
                           & $34.8 \pm 0.3 \pm 1.0$   \\
      Tagging power   [\%] & $0.68 \pm 0.03 \pm 0.05$
                           & $0.30 \pm 0.02 \pm 0.04$ \\ \hline
    \end{tabular}
  \end{center}
  \caption{Performance of flavour tagging algorithm in CMS analysis; errors 
    are statistical only.}
  \label{tab:cmsft}
\end{table}

%\begin{table}[htb]
%  \begin{center}
%    \begin{tabular}{l|ll}
%      \multicolumn{1}{c|}{} &
%      \multicolumn{1}{c}{ATLAS} &
%      \multicolumn{1}{c}{CMS} \\ \hline
%      Tagging efficiency $\epsilon_\tsf (\%)$
%      & $32.1 \pm 0.01$ & $7.67 \pm 0.04$ \\
%      Mistag fraction $\omega (\%)$
%      & $21.3 \pm 0.08$ & $32.2 \pm 0.3$ \\
%      Tagging power $P_\tsf (\%)$
%      & $1.45 \pm 0.05$ & $0.97 \pm 0.03$ \\ \hline
%    \end{tabular}
%  \end{center}
%  \caption{Performance of flavour tagging algorithm; errors 
%    are statistical only.}
%  \label{tab:bjpft}
%\end{table}

Efficiency and background have been estimated using simulation; background 
is made essentially by decay of other $B$ hadrons as $\bd$~, $\bp$~, 
$\Lambda_b$ and $B_c$. 
Efficiency have been estimated comparing the fully simulated signal sample, 
including detector response, with generator level information and computing 
the efficiency as a function of the three angles. 

\begin{figure}[htb]
  \centering
  \includegraphics[height=45mm]{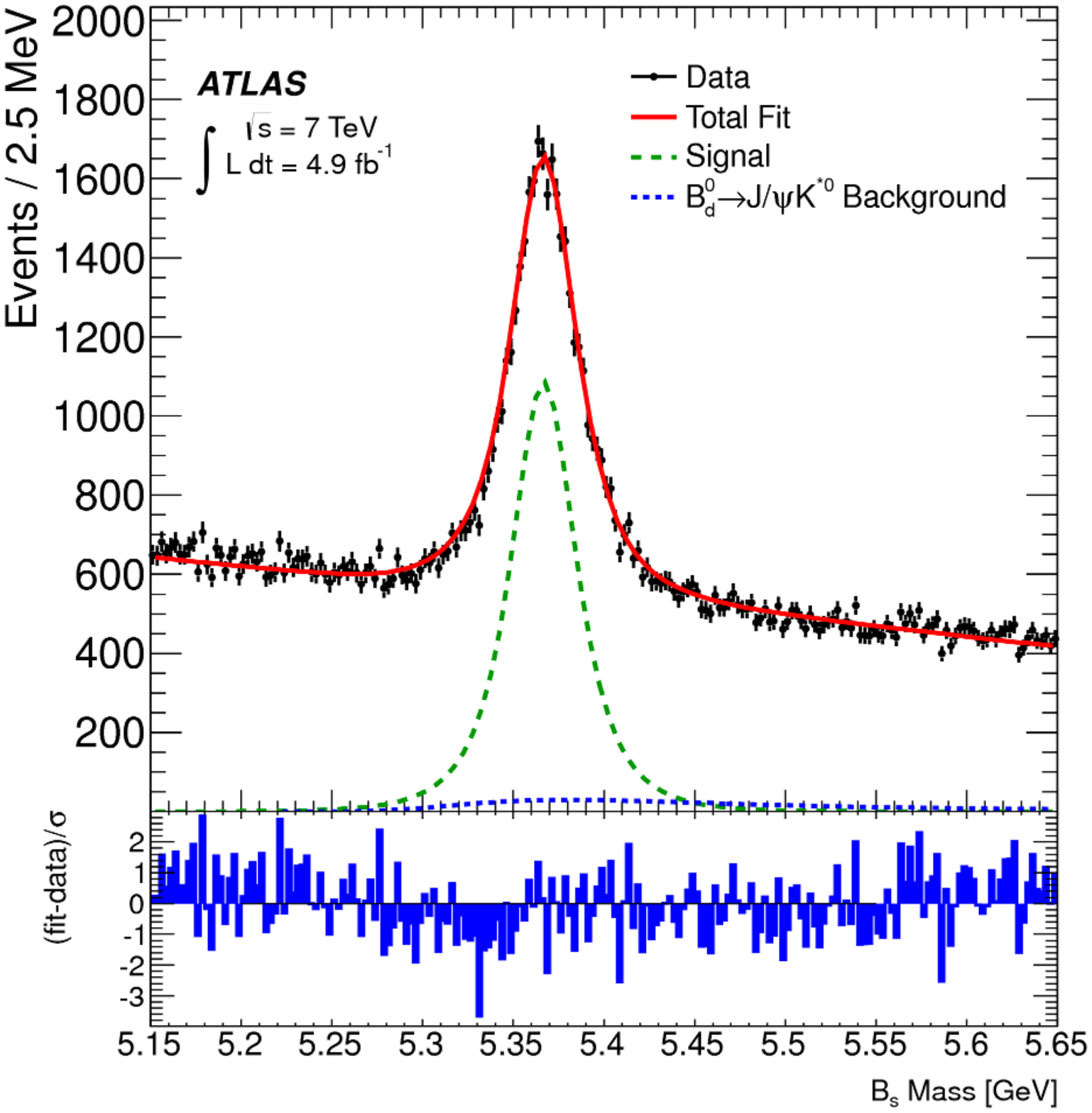}
  \rule{8mm}{0pt}
  \includegraphics[height=45mm]{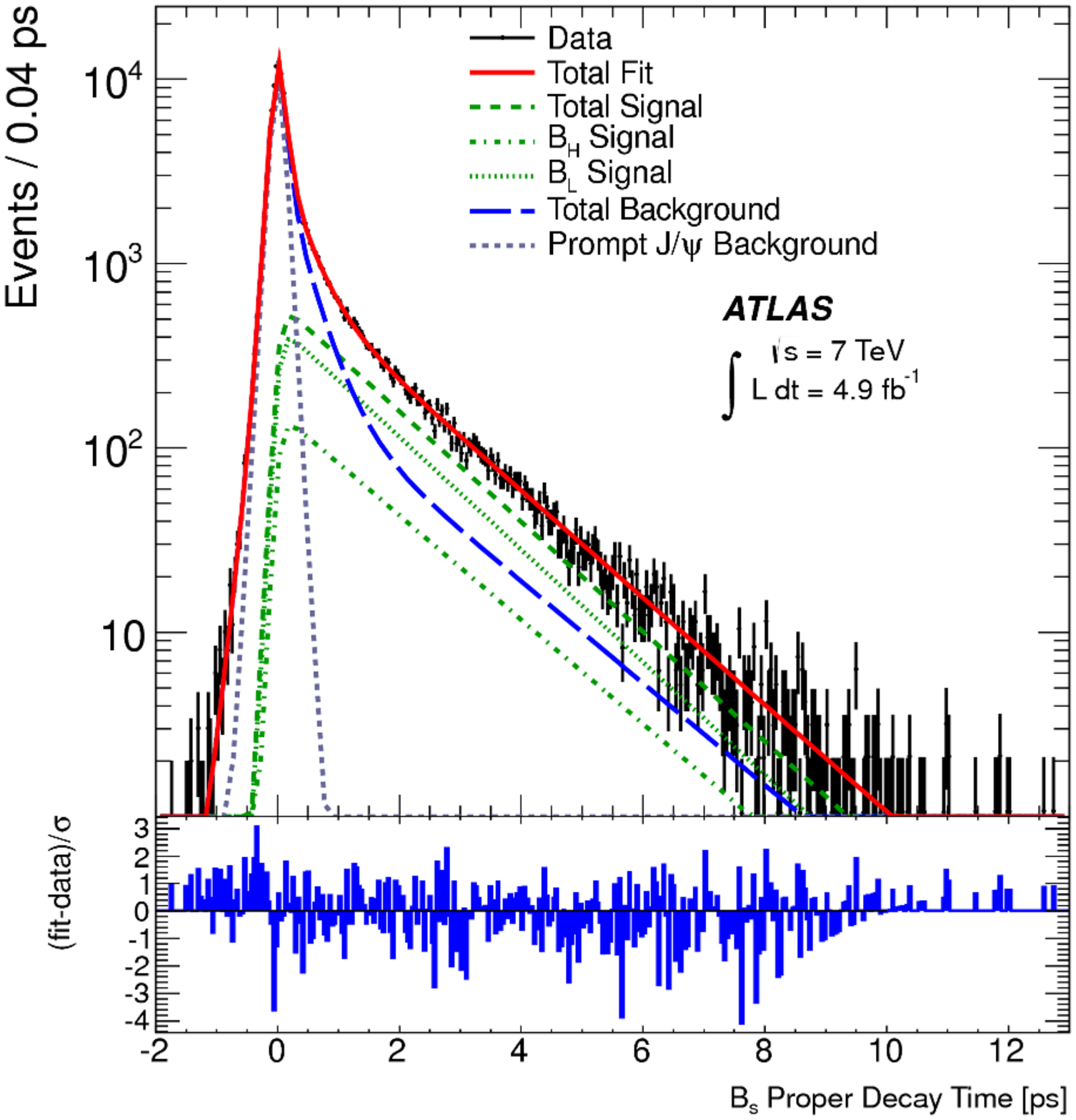}
  \caption{Invariant mass (left) and proper decay length (right) distributions
           in ATLAS fit.}
  \label{fig:bjpfa}
\end{figure}

\begin{figure}[htb]
  \centering
  \includegraphics[height=42mm]{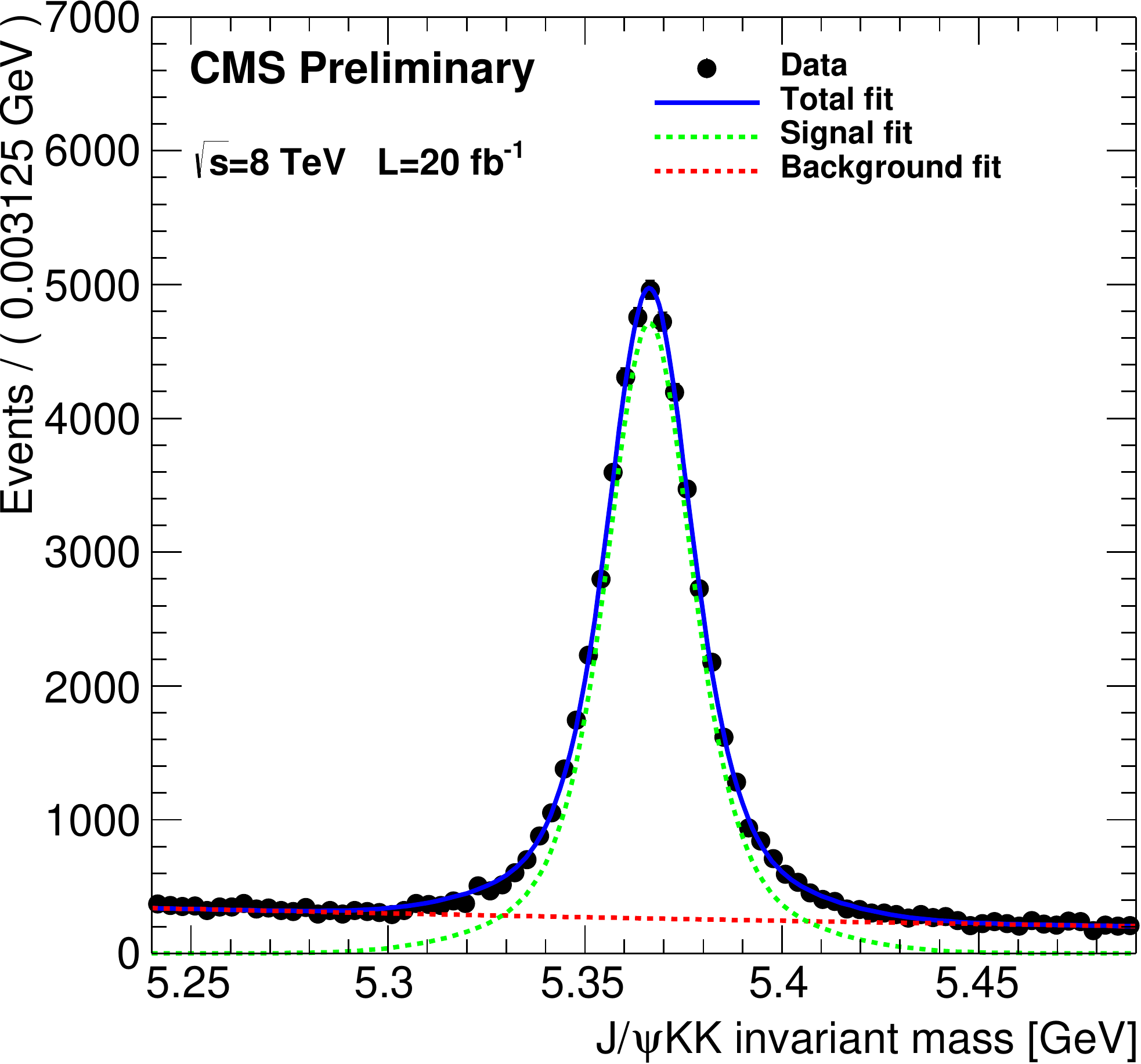}
  \rule{8mm}{0pt}
  \includegraphics[height=42mm]{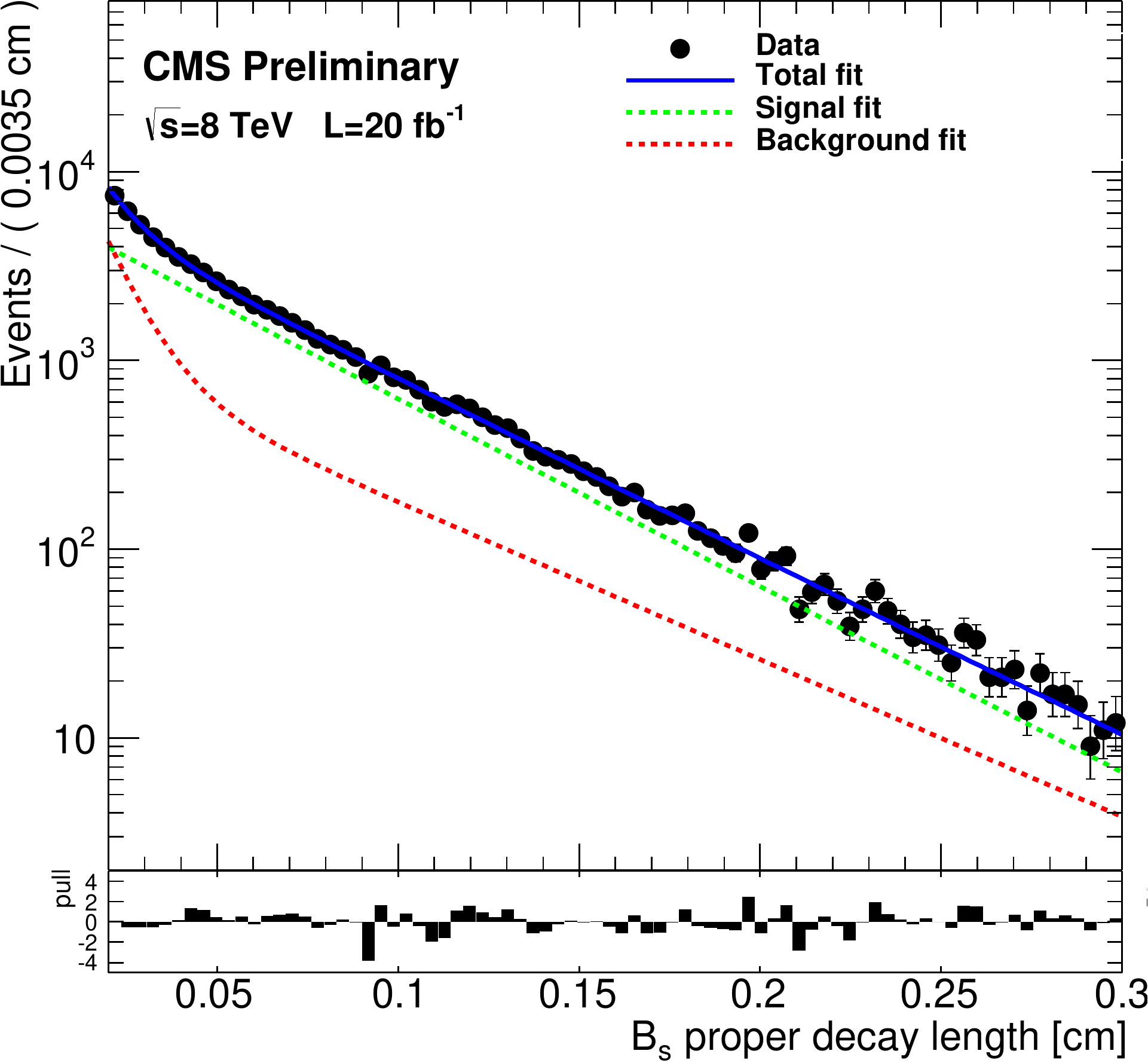}
  \caption{Invariant mass (left) and proper decay time (right) distributions
           in CMS fit.}
  \label{fig:bjpfc}
\end{figure}

An unbinned maximum likelihood fit has been performed, including per-event 
resolution and tagging probability terms. The likelihood function is a sum 
of signal and background terms, with model functions for the three angles 
and the invariant mass, decay time and tagging information.
The fitted distributions of invariant mass and proper time, or length, are 
shown in fig.\ref{fig:bjpfa} for ATLAS and .fig.\ref{fig:bjpfc} for CMS. 

Systematic uncertainties arise from efficiency and background estimation, 
from resolution and model functions. Potential biases have been estimated 
by pseudo-esperiments and included in the systematic error. 

\section{Results and expectations}

\begin{figure}[htb]
  \centering
  \includegraphics[height=138pt]{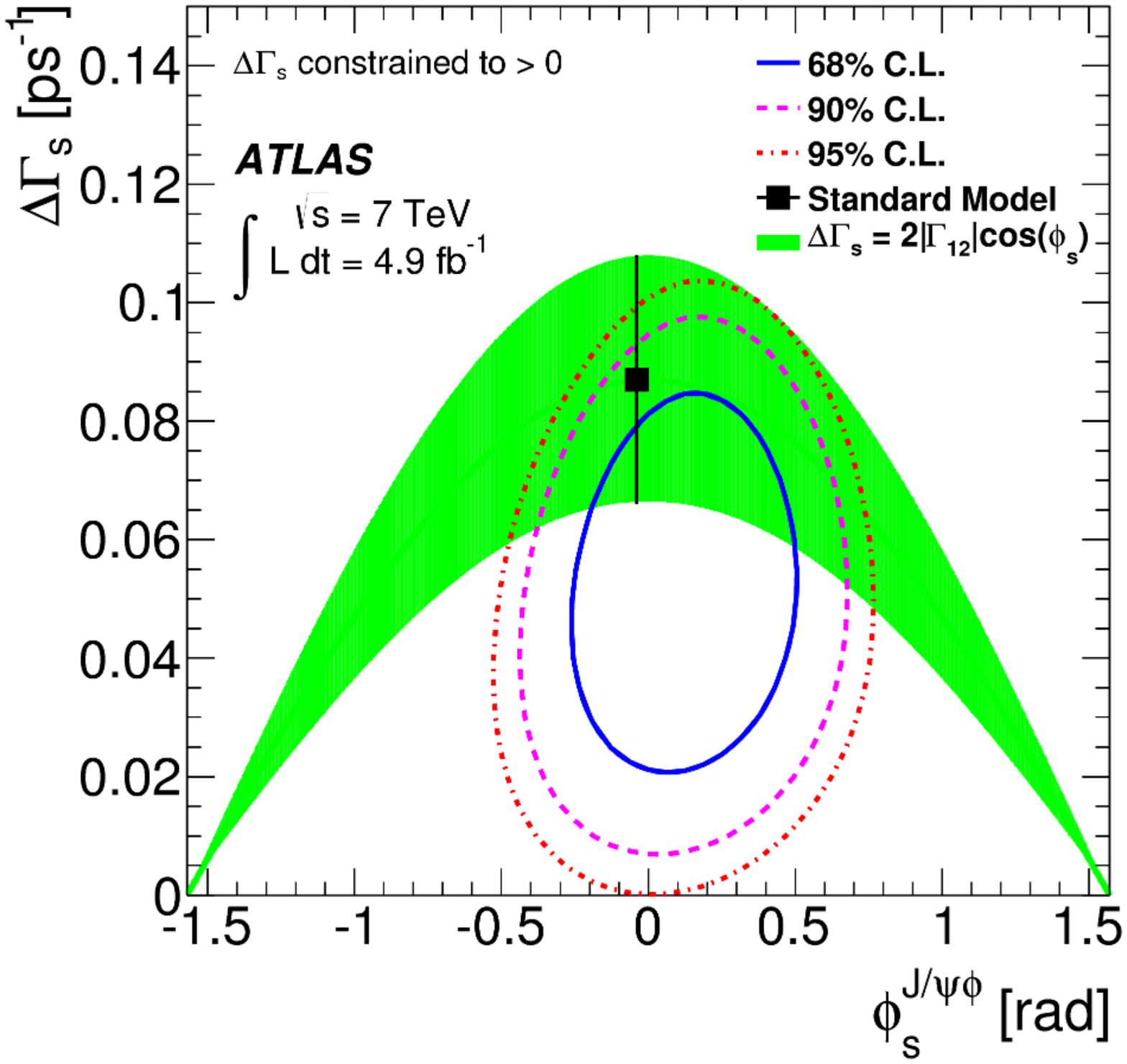}
  \rule{8mm}{0pt}
  \includegraphics[height=144pt]{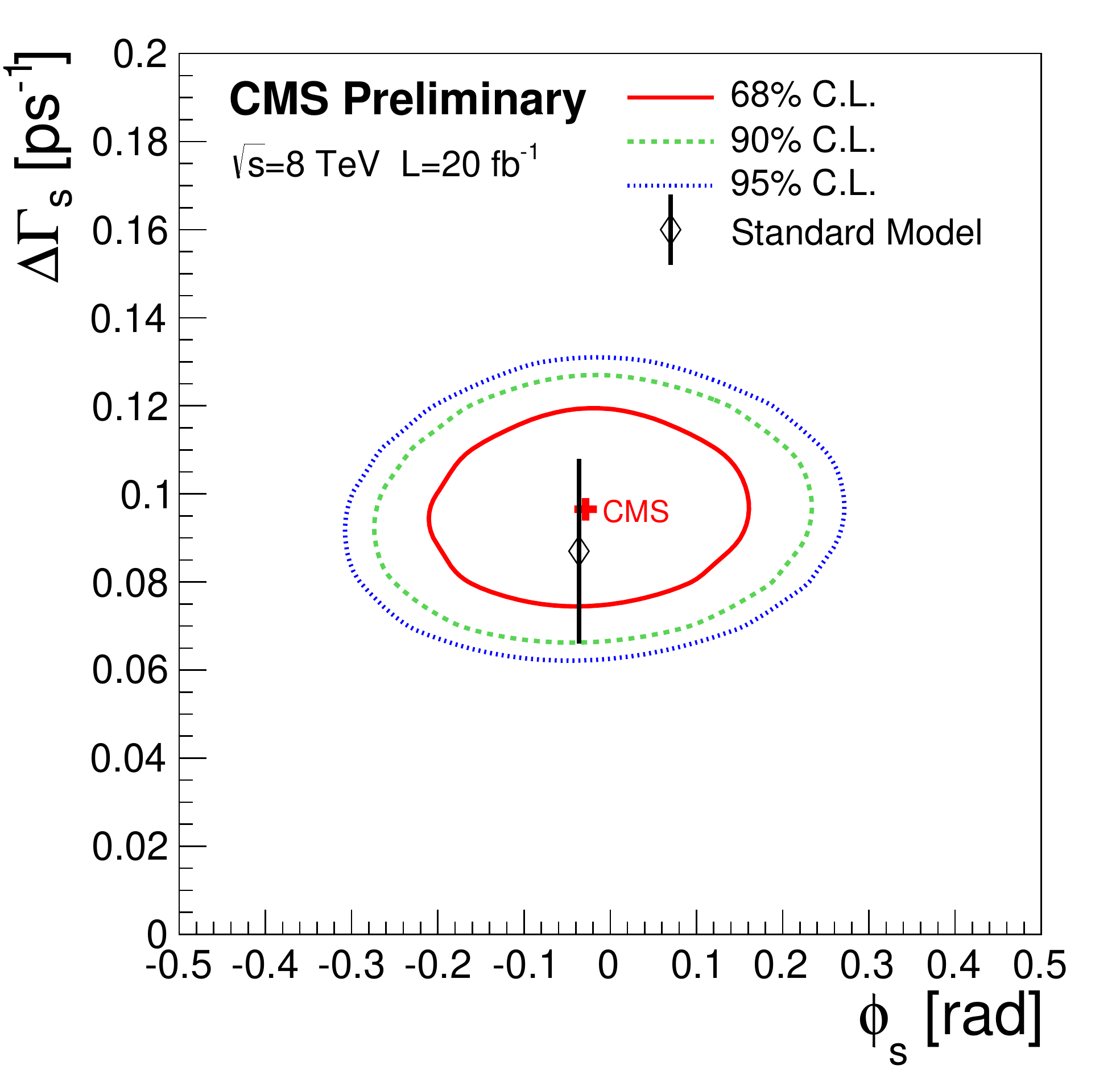}
  \caption{Likelihood contours in the $\phs-\dgs$ plane in ATLAS (left) 
           and CMS (right) measurements.}
  \label{fig:bjpfr}
\end{figure}

The results of the 
fit are shown in tab.\ref{tab:bjpfr} and fig.\ref{fig:bjpfr}; 
both experiments found 
a result in agreement with the prediction, but a significant test will 
require a smaller uncertainty, so further investigations are required.

\begin{table}[htb]
  \begin{center}
    \begin{tabular}{l|cc}
      \multicolumn{1}{c|}{} & ATLAS & CMS \\ \hline
      $\phs [\mathrm{rad}]$
      & $0.12  \pm 0.25  \pm 0.05 $ & $-0.03 \pm 0.11  \pm 0.03 $ \\
      $\dgs [\ips]$
%      \rule{0pt}{11pt}
      & $0.053 \pm 0.021 \pm 0.010$ & $0.096 \pm 0.014 \pm 0.007$ \\
      \hline
    \end{tabular}
 \end{center}
  \caption{Final results of CP violating phase and decay width difference 
    measurements.}
  \label{tab:bjpfr}
\end{table}

\begin{figure}[htb]
  \centering
  \includegraphics[height=45mm]{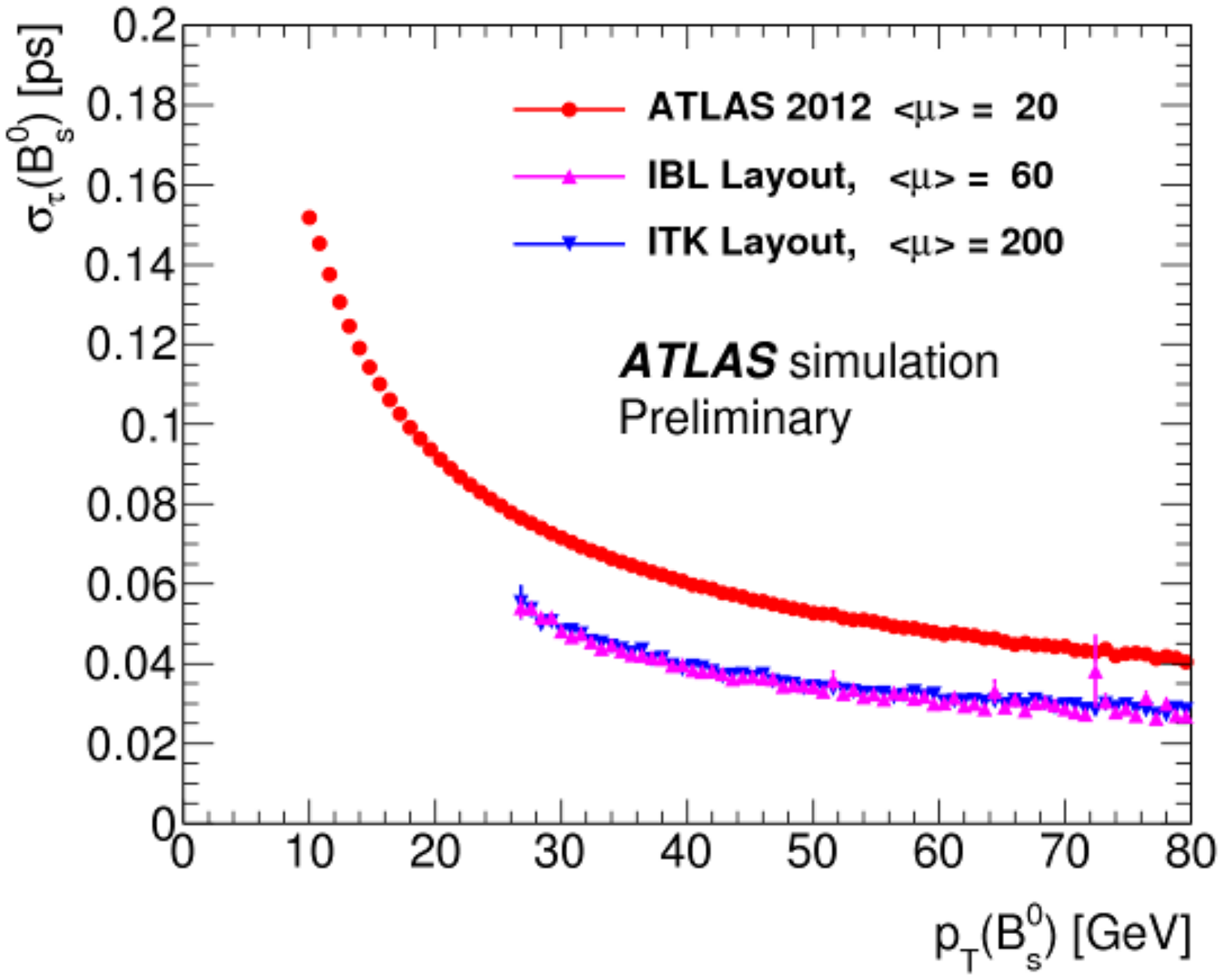}
  \rule{8mm}{0pt}
  \includegraphics[height=45mm]{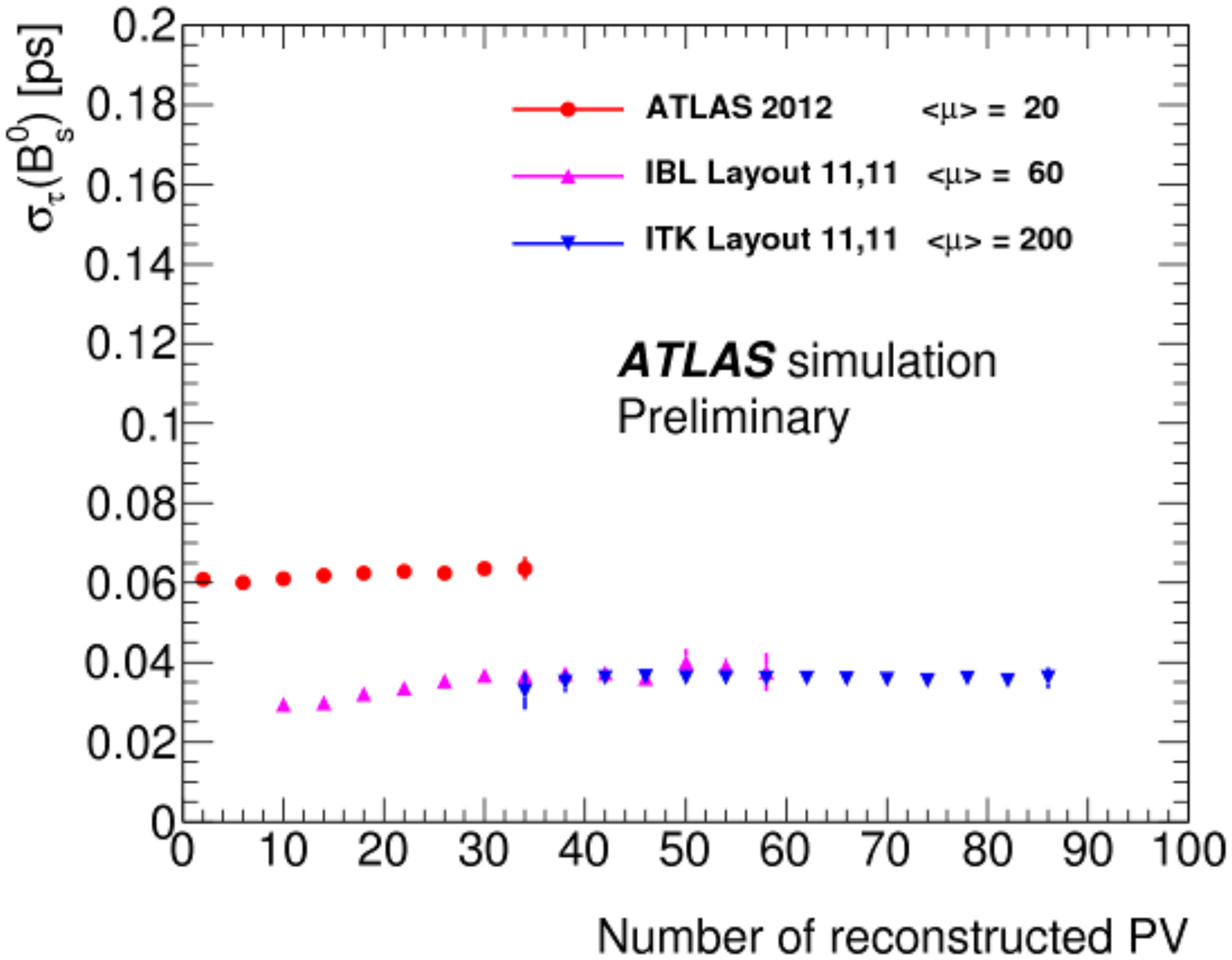}
  \caption{Proper decay time resolution in ATLAS against the 
    $\bs$ transverse momentum (left) and the number of 
    reconstructed primary vertices (right) in simulated $\bsjphi$ events.}
  \label{fig:bjpnr}
\end{figure}

ATLAS made a study to estimate the measurement potential with new runs; 
more data will be available with the new run and increased 
luminosity, but this will correspond to a more difficult environment. 
ATLAS will have an improved pixel tracker, with a fourth layer, for Run2, 
and a new tracker with reduced pixel size for HL-LHC. 
The need to stay inside a necessarily limited trigger bandwidth will require 
harder cuts on muon $p_T$; two possible cuts have been considered, at 
$6~\gev$ for Phase-1 and $11~\gev$ for Phase-2~\cite{ref:bjpnr}.
The upgraded tracker will allow a better vertex reconstruction and an 
improvement of 30\% in proper decay time resolution, as shown in 
fig.\ref{fig:bjpnr} (left).
In principle this could be affected by the higher pileup, but a dedicated 
study showed that even in the hypothesis that the number 
of interaction could 
reach $N = 200$, no significant effect appears visible, as shown in 
fig.\ref{fig:bjpnr} (right).

Estimating the signal yields applying the harder muon $p_T$ cuts to 
2012 data and rescaling with efficiencies and luminosities 
the results shown in tab.\ref{tab:bjpnr} are obtained. 
\begin{table}[htb]
  \begin{center}
    \begin{tabular}{c|cc}
%      ${\cal L}(\ifb)$ &
%      $p_{T\mu}$ cut $[\gev]$ &
%      $\sigma(\phi_s)$(stat)$[\rad]$ \\ \hline
      \raisebox{0pt}[11pt][5pt]{${\cal L}(\ifb)$} &
      \raisebox{0pt}[11pt][5pt]{$p_{T\mu}$ cut $[\gev]$} &
      \raisebox{0pt}[11pt][5pt]{$\sigma(\phi_s)$(stat)$[\rad]$} \\ \hline
       100 &  6 & 0.054 \\
       100 & 11 & 0.10  \\
       250 & 11 & 0.064 \\
       3000 & 11 & 0.022 \\ \hline
    \end{tabular}
  \end{center}
  \caption{Estimated ATLAS statistical precisions of $\phs$ measurement 
    for considered LHC periods.}
  \label{tab:bjpnr}
\end{table}

\section{Conclusions}

The measurement of the CP violating phase $\phi_s$ can give hints or
constraints of new physics beyond the standard model.

ATLAS and CMS performed a measurement by mean of an angular analysis of 
the decay $\bsjphi$ using an initial flavour tagging to increase sensitivity.
Results are compatible with previous result and SM expectations, but the 
uncertainty is currently much bigger than the theoretical error. 
More stringent tests will be obtained with more precise measurement to be 
done in the future LHC runs.

%\Acknowledgements
%xxxxxxxx yyyy zzzz.

\urlstyle{same}

\newcommand{\cmscoll}{CMS Collaboration}
\newcommand{\cmslhcb}{CMS and LHCb Collaborations}
\newcommand{\atlcoll}{ATLAS Collaboration}
\newcommand{\lhbcoll}{LHCb Collaboration}

\newcommand{\eal}{{\it et al.}}

\newcommand{\tit}{}

\newcommand{\tpr}{\tit,}

\end{document}